\documentclass[
  aps,
  pra,
  amsmath,
  amssymb,
  reprint,
  superscriptaddress,
  floatfix,
  longbibliography,
  showkeys
]{revtex4-2}

\usepackage{graphicx}
\usepackage{bm}
\usepackage[hidelinks]{hyperref}

\def\tsc#1{\csdef{#1}{\textsc{\lowercase{#1}}\xspace}}
\tsc{WGM}
\tsc{QE}
\tsc{EP}
\tsc{PMS}
\tsc{BEC}
\tsc{DE}

\begin{document}
\title{Position- and Momentum-Space Quantum Information Measures of the Double-Morse Oscillator}   

\author{F. Chogle}
\affiliation{College of Computing and Mathematical Sciences, Department of Applied Mathematics and Sciences, Khalifa University of Science and Technology, Abu Dhabi 127788, United Arab Emirates}

\author{Ernesto Damiani}
\affiliation{KU Research Center for Advanced Intelligent Systems (AIS), Khalifa University of Science and Technology, 127788, Abu Dhabi, United Arab Emirates}
\affiliation{Dipartimento di Informatica, Universit{\`a} degli Studi di Milano, Via Giovanni Celoria 18}

\author{B. Teklu}
\email{berihu.gebrehiwot@ku.ac.ae}
\affiliation{College of Computing and Mathematical Sciences, Department of Applied Mathematics and Sciences, Khalifa University of Science and Technology, Abu Dhabi 127788, United Arab Emirates}
\affiliation{KU Research Center for Advanced Intelligent Systems (AIS), Khalifa University of Science and Technology, 127788, Abu Dhabi, United Arab Emirates}

\date{August 26, 2026}

\begin{abstract}
We investigate the quantum-information properties of a particle confined by the double Morse potential in position and momentum spaces.
The quasi-exact solvability of the model gives analytical expressions for the first two bound states, allowing the corresponding probability densities to be analyzed directly.
Shannon entropy, Onicescu energy, Fisher information, statistical complexity, and Fisher-Shannon products are evaluated as functions of the parameter $A$, which controls the transition from a well-separated double well to a merged single-well profile.
The position distribution is more delocalized and structurally complex when the double-well character is pronounced, whereas the momentum distribution exhibits the complementary trend.
As the wells merge, the ground-state Fisher--Shannon product approaches its Gaussian reference value, whereas the excited state retains stronger non-Gaussian structure.
\end{abstract}

\maketitle
\section{Introduction}\label{sec:intro}
Information-theoretic measures provide a complementary description of quantum systems beyond expectation values of conventional observables. While moments of position, momentum, and energy characterize selected averages of a state, information measures quantify the full spatial and momentum probability distributions and can therefore reveal localization, delocalization, oscillatory structure, and sensitivity to changes in a control parameter \cite{mukherjee2019some,mukherjee2020analysis,ojha2025quantum,edet2021shannon,olendski2018quantum}. This viewpoint has been fruitfully applied to atoms, confined oscillators, quantum wells and dots, and, more recently, waveguide geometries with mixed boundary conditions \cite{chogle2026quantuminformationdirichletneumann}. In the present work we use this framework to study a double Morse oscillator, whose tunable double-well structure makes it a useful model for connecting spatial localization, anharmonicity, and quantum-state complexity.

The first measure considered here is the Shannon information entropy, introduced as a measure of uncertainty in a probability distribution \cite{shannon1948mathematical}. In position and momentum spaces it is defined as
\begin{subequations}
\begin{align}
    \mathcal{S}_\rho & = -\int \rho(x)\ln \rho(x)\,dx, \label{srho_gen} \\
    \mathcal{S}_\gamma & = -\int \gamma(k)\ln \gamma(k)\,dk, \label{sgamma_gen}
\end{align}
\end{subequations}
where $\rho(x)=|\psi(x)|^2$ and $\gamma(k)=|\phi(k)|^2$ are the normalized position and momentum probability densities, respectively. The entropic uncertainty relation
\begin{equation}
    \mathcal{S}_\rho + \mathcal{S}_\gamma \geq 1+\ln\pi
    \label{st_inequality}
\end{equation}
follows from Fourier-analysis bounds and provides a distribution-level formulation of quantum uncertainty \cite{beckner1975inequalities,bialynicki1975uncertainty}. Shannon entropy has been used to characterize many analytically and numerically solvable potentials, including the harmonic oscillator, P\"{o}schl--Teller and squared-tangent wells, Morse, and Rosen--Morse potentials \cite{gadre1985some,sun2013quantum,dong2014squaredtangent,abdelmonem2017information,sun2013quantumrosen,sun2013quantum3}. A localized density has lower entropy, whereas a spread density has larger entropy; in conjugate space these trends are constrained by Eq.~(\ref{st_inequality}).

A complementary global measure to Shannon entropy is the Onicescu energy \cite{onicescu1966}, which quantifies the disequilibrium or concentration of a probability density. It is defined as
\begin{subequations}
\begin{align}
    \mathcal{O}_\rho & = \int \rho^2(x)\,dx, \label{orho_gen} \\
    \mathcal{O}_\gamma & = \int \gamma^2(k)\,dk. \label{ogamma_gen}
\end{align}
\end{subequations}
Unlike Shannon entropy, which grows with delocalization, the Onicescu energy grows when a density becomes more concentrated. For quantum states one obtains the reference product
\begin{equation}
    \mathcal{O}_\rho \mathcal{O}_\gamma \leq \frac{1}{2\pi},
    \label{oni_inequality}
\end{equation}
For a Gaussian wave packet the product equals $1/(2\pi)$. The inequality is often used as a reference for broad classes of states, but it is not a universal uncertainty relation and counterexamples are known \cite{ghosal2016information,shafeekali2023quantum}. Accordingly, throughout this paper $1/(2\pi)$ is interpreted as a Gaussian benchmark, not as an inviolable bound. Combining entropy and disequilibrium leads to the Catalan-Garay-Lopez-Ruiz statistical complexity
\begin{equation}
    \mathrm{CGL} = e^\mathcal{S} \mathcal{O},
    \label{eqn_cgl}
\end{equation}
which captures the simultaneous roles of spreading and structural organization in a density distribution \cite{lopezruiz1995statistical,catalan2002features,lopez2009rigorous}.

Shannon entropy and Onicescu energy are global descriptors and do not directly quantify local gradients of a probability density. This local information is provided by Fisher information \cite{fisher1925theory},
\begin{subequations}
\begin{align}
    \mathcal{I}_\rho & = \int \frac{1}{\rho(x)}\left(\frac{\partial \rho(x)}{\partial x}\right)^2\,dx, \label{irho_gen} \\
    \mathcal{I}_\gamma & = \int \frac{1}{\gamma(k)}\left(\frac{\partial \gamma(k)}{\partial k}\right)^2\,dk. \label{igamma_gen}
\end{align}
\end{subequations}
In estimation theory Fisher information bounds the variance of an unbiased estimator through the Cram\'{e}r--Rao relation, $\mathrm{Var}(\hat{A})\geq 1/I_A$, and in quantum physics it plays a central role in quantum metrology and sensing \cite{paris2009quantum,adani2024criticalmetrology}. In conjugate spaces the Fisher-information product is often compared with
\begin{equation}
    \mathcal{I}_\rho \mathcal{I}_\gamma \geq 4,
    \label{fisher_ineq}
\end{equation}
although counterexamples show that this relation must be used with care \cite{romera2006uncertainty,choi2011information,aguiar2015shannon,aguiar2015fisher,aguiar2016entropy}. Fisher information can also be combined with the entropy power to form the Fisher-Shannon product
\begin{equation}
    \mathcal{P}=\mathcal{I}\mathcal{J}, \qquad
    \mathcal{J}=\frac{e^{2\mathcal{S}}}{2\pi e},
    \label{fisher-shannon-product}
\end{equation}
which equals unity for a Gaussian distribution and is useful for diagnosing non-Gaussianity and structural complexity \cite{romera2004,dembo,angulo2008fishershannon,AQUINO20132062,IKOT2020103150}. 

The double Morse potential is especially interesting in this context because its morphology can be tuned continuously from two well-separated wells to a merged single-well profile. Such tuning changes the localization of the bound states and, at the same time, modifies the anharmonic and non-Gaussian character of the system. In a closely related double-well setting, Shannon and Fisher measures, disequilibrium, and LMC complexity were shown to provide a sensitive description of barrier-mediated tunneling in a double-square-well model of ammonia \cite{tserkis2014tunneling}. Recent work on nonlinear quantum oscillators has shown that nonlinearity can be quantified through ground-state distances and non-Gaussianity \cite{paris2014quantifyingnonlinearity}, and that nonlinearity can act as a resource for nonclassicality in nanomechanical and nonlinear optical systems \cite{teklu2015nonlinearitynanomechanical,candeloro2021quantumprobes,asjad2023jointquantumestimation}.  Recent studies have addressed complementary aspects of the same double-Morse model. Ref.~\cite{Chogle2026} treated nonlinearity and nonclassicality as quantum resources and investigated the metrological estimation of the control parameters $\alpha$ and $A$. Ref.~\cite{Chogle2026Quasiprobability} instead developed an exact phase-space description of the lowest quasi-exact ground state, deriving its Wigner function, Weyl characteristic function, and full Cahill--Glauber $s$-ordered quasiprobability hierarchy. The present work complements these studies by characterizing the full position- and momentum-space probability densities through information-theoretic measures and tracking the redistribution of localization, uncertainty, and complexity across the double-well--to--single-well crossover.

Motivated by these developments, we investigate the position- and momentum-space information measures of the two analytically accessible states of the quasi-exactly solvable double Morse oscillator \cite{KONWENT1986467,ushveridze2017quasi}. Previous studies examined the optical response of double-Morse quantum wells \cite{kasapoglu2019optical} and the nonlinearity, nonclassicality, and metrological properties of the double-Morse oscillator \cite{Chogle2026}; however, a systematic information-theoretic analysis of the model remains missing. Here we exploit the quasi-exact solvability of the system to compute Shannon entropy, Onicescu energy, Fisher information, statistical complexity, and Fisher-Shannon products as functions of the dimensionless parameter $A$. This analysis clarifies how the merging of the two wells redistributes uncertainty between position and momentum spaces and how the ground state approaches Gaussian-like behavior in the single-well regime. 
\section{Model and Formulation}\label{sec:model}
The interaction between two atoms in a diatomic molecule is commonly described by the single Morse potential
\begin{equation}
    V_\mathrm{SM}(x) = D(e^{-2\alpha(x-x_0)} - 2e^{-\alpha(x-x_0)}),
    \label{single_morse}
\end{equation}
where $D$ is the bond-dissociation energy, $\alpha$ controls the well skewness, and the well minimum is located at $x_0$.
The double Morse potential is obtained by adding two single Morse potentials centered at $\pm x_0$ and facing each other.
After dropping an additive constant, which shifts the zero of energy without changing the physics, the potential becomes $V_\mathrm{DM}=D(A\cosh(\alpha x)-1)^2$ with $A = 2e^{-\alpha x_0}$.
The double-well structure occurs for $0<A<1$.

Consider a particle of mass $m$ under the influence of the double Morse potential.
Its position wavefunction $\psi(x)$ is found by solving the time-independent Schr\"odinger equation 
\begin{equation}
    -\frac{\hbar^2}{2m}\frac{\partial^2 \psi(x)}{\partial x^2} + D(A\cosh(\alpha x)-1)^2 \psi(x) = E\psi(x)
    \label{eqn1}
\end{equation} 
where $E$ is the energy of the particle in its corresponding state $\psi(x)$. Methodology of \cite{KONWENT1986467} is used to solve equation (\ref{eqn1}).
Introducing the dimensionless transformations,
\begin{equation}
    \alpha x = 2y,\ \mu^2 = \frac{8mD}{\hbar^2\alpha^2},\ \epsilon = \frac{8mE}{\hbar^2\alpha^2}
    \label{dimensionless_var}
\end{equation}
yields the following,
\begin{equation}
    \frac{d^2\psi}{dy^2} + [\epsilon - \mu^2(A\cosh(2y)-1)^2]\psi = 0.
    \label{eqn2}
\end{equation}
\begin{figure}
    \centering
    \includegraphics[width=0.8\columnwidth]{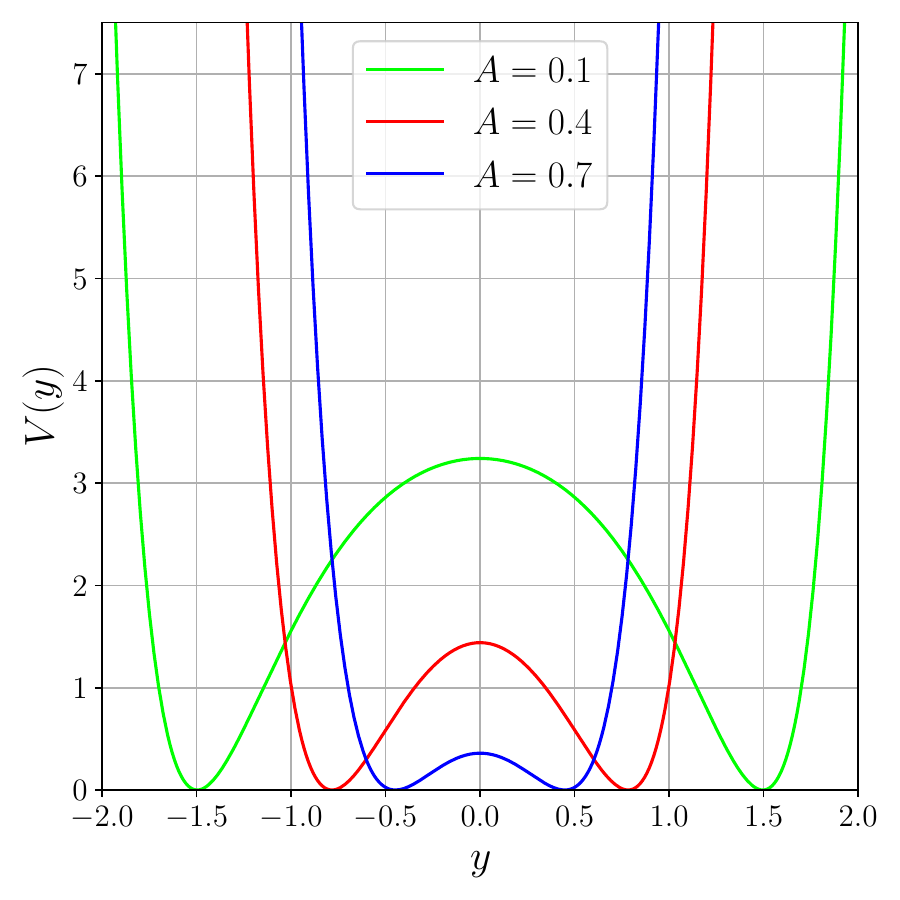}
    \caption{Double Morse potential for $A=0.1,0.4,0.7$ and $\mu=2$.}
    \label{fig1}
\end{figure}
Figure~\ref{fig1} shows the plot of the double Morse potential in dimensionless units, i.e., $V(y) = \mu^2(A\cosh(2y)-1)^2$. The influence of the 
well's structure is governed by $A$. For small values of $A$, the potential shows two distinct wells and as $A\to1^{-}$, the two wells merge into one. 

Since $A$ governs the transition from a double-well potential to a merged single-well profile, our analysis focuses on $A\in(0,1)$.
This parametrization is useful because $A$ incorporates the combined effects of the skewness parameter $\alpha$ and the inter-well separation $x_0$.
In Ref.~\cite{Chogle2026}, the quantumness of the system was attributed to changes in $\alpha$ while $x_0$ was held fixed.
Here we emphasize that $x_0$ is also important because it controls the separation of the two wells and, consequently, the localization of the particle.
Thus, varying $A$ provides a compact way to study the information-theoretic properties of the system.

The asymptotic solution of equation (\ref{eqn2}) for $y \to \pm \infty$ is given as $\psi_a(y) = \exp\left(-\frac{\mu A}{2}\cosh(2y)\right)$.
Physically, this describes the semi-classical or WKB approximation of the wavefunction. The quantum description of the wavefunction is found by 
assuming that the solution is of the form $\psi(y) = \psi_a(y)\phi(y)$ where the unknown function $\phi(y)$ satisfies
\begin{equation}
    \frac{d^2\phi}{dy^2} - 2\mu A\sinh2y\frac{d\phi}{dy} + (\tilde{\epsilon}+2\mu A(\mu-1)\cosh2y)\phi = 0
    \label{eqn4}
\end{equation}
where $\tilde{\epsilon} = \epsilon-\mu^2(1+A^2)$. Analytical solution of the bound state is obtained when $\phi(y)$ is a polynomial of hyperbolic functions.
This is achieved when $\mu = n+1$. As a result, the system has $n+1$ discrete levels and classified as quasi-exactly solvable.\\

\subsection{Experimental realizations and parameter interpretation}\label{sec:experimental}
This section clarifies the possible experimental meaning of the dimensionless quantities used below. In contrast to Ref.~\cite{Chogle2026}, which focuses on metrological estimation of the inverse barrier-width parameter $\alpha$ and the reparametrized variable $A=2e^{-\alpha x_0}$, here we interpret the information-theoretic trends of Sec.~\ref{sec:qim}: $A$ is a compact shape parameter for the probability densities, while $\mu$ fixes the quasi-exactly solvable sector.
Consequently, the curves reported below should be read as information-theoretic diagnostics of a calibrated effective potential, rather than as a full device-specific estimation protocol.

In physical units, $D$ fixes the energy scale, $\alpha$ is an inverse length that controls the steepness of each Morse branch, and $x_0$ is the half-separation of the two parent Morse wells.
The experimentally accessible shape parameter is
\begin{equation}
    A=2e^{-\alpha x_0},
    \label{eq:A_parameter_interpretation}
\end{equation}
so changing a laser intensity, electrode voltage, strain field, bond geometry, or heterostructure profile can modify the effective value of $A$ either by changing the separation scale $x_0$, the inverse length $\alpha$, or both.
For $0<A<1$, the minima of $V_\mathrm{DM}(x)$ occur at
\begin{equation}
    x_\mathrm{min}=\pm\frac{1}{\alpha}\operatorname{arcosh}\left(\frac{1}{A}\right),
    \label{eq:xmin_exp}
\end{equation}
and the central barrier height is
\begin{equation}
    V_\mathrm{DM}(0)=D(1-A)^2.
    \label{eq:barrier_exp}
\end{equation}
The local harmonic curvature at either minimum is
$V_\mathrm{DM}''(x_\mathrm{min})=2D\alpha^2(1-A^2)$, which yields the local oscillation frequency $\omega_\mathrm{loc}=\alpha\sqrt{2D(1-A^2)/m}$. Together, these expressions provide experimentally accessible calibration targets: measurements of the well separation, barrier height, local oscillation frequency, and tunneling splitting can be used to fit an effective double-Morse potential before the corresponding information measures are evaluated.

An important caveat is that $\mu^2=8mD/(\hbar^2\alpha^2)$ depends on both $D$ and $\alpha$. Consequently, varying $\alpha$ while keeping $D$ fixed changes both $A$ and $\mu$. The fixed-$\mu$ curves obtained in the present quasi-exact analysis, therefore, represent either a protocol in which $D/\alpha^2$ is co-tuned to remain constant or a local benchmark for a device whose two lowest states are well described by the same effective value of $\mu$.

Several platforms can realize or approximate this situation.
In ultracold atoms, optical double wells have already been used for condensate interferometry and bosonic Josephson dynamics, while painted optical potentials and optical superlattices provide highly tunable one-dimensional or quasi-one-dimensional landscapes \cite{shin2004atominterferometry,albiez2005directobservation,henderson2009painting,chalopin2025opticalsuperlattice}.
In such settings, $A$ may be calibrated from the optical barrier height and the well separation, the position density can be obtained from in-situ or reconstructed density profiles, and the momentum density can be obtained from time-of-flight expansion. Operationally, a measured, background-subtracted profile is normalized to probabilities $p_i=N_i/\sum_jN_j$ in detector bins of width $\Delta y$; the differential entropy is then estimated as $\mathcal{S}_{\rho}\simeq-\sum_i p_i\ln(p_i/\Delta y)$. The same procedure applied to calibrated time-of-flight momentum bins yields $\mathcal{S}_{\gamma}$. Finite resolution, limited field of view, and counting noise should be propagated by detector-response deconvolution or resampling; Fisher information is more noise-sensitive because it involves derivatives, whereas Shannon entropy and Onicescu energy can be evaluated directly from the normalized histogram.
In semiconductor and laser-dressed quantum wells, double-Morse-type confinement has been used to analyze optical absorption and nonlinear optical response under changes of structure and laser parameters \cite{kasapoglu2019optical}.
Segmented trapped-ion devices provide another route, since electrode voltages can reshape axial confinement and experiments have demonstrated both controlled ion separation and measurable anharmonic trap effects \cite{home2011normalmodes,bowler2012coherentdiabatic}.
Finally, in hydrogen-bonded and proton-transfer systems, two-Morse or double-minimum potentials arise naturally; NMR and related spectroscopic data have been used to infer microscopic Morse-like proton potentials in KDP-type materials \cite{mashiyama2004dynamical,mashiyama2007quasiharmonic,kim2020microscopic}.
These examples show that the model should not be interpreted as a single hardware prescription.
Rather, it supplies a transferable effective-potential language in which the information measures quantify how localization, momentum spreading, and non-Gaussian structure respond to experimentally calibrated changes of double-well shape.

\section{Wavefunctions}\label{sec:wave}
For $n = 1$, the system supports two quasi-exact levels: the ground state and the first excited state.
The position wavefunctions (ground and excited states) of the system are given by 
\begin{align}
    \psi_0(y) &= \sqrt{\frac{2}{K_1(2A)+K_0(2A)}} \cosh(y)e^{-A\cosh(2y)} \label{eqn5} \\
    \psi_1(y) &= \sqrt{\frac{2}{K_1(2A)-K_0(2A)}} \sinh(y)e^{-A\cosh(2y)} \label{eqn6}
\end{align}
and their corresponding energies are 
\begin{align}
    \epsilon_0 &= -1 -4A + 4(1+A^2) \label{eqn7} \\
    \epsilon_1 &= -1 +4A + 4(1+A^2) \label{eqn8}.
\end{align}
The function $K_{\nu}(z)$ is the $\nu^{\text{th}}$ order modified Bessel function of the second kind \cite{abramowitz1948handbook}.

\begin{figure}
    \centering
    \includegraphics[width=\columnwidth]{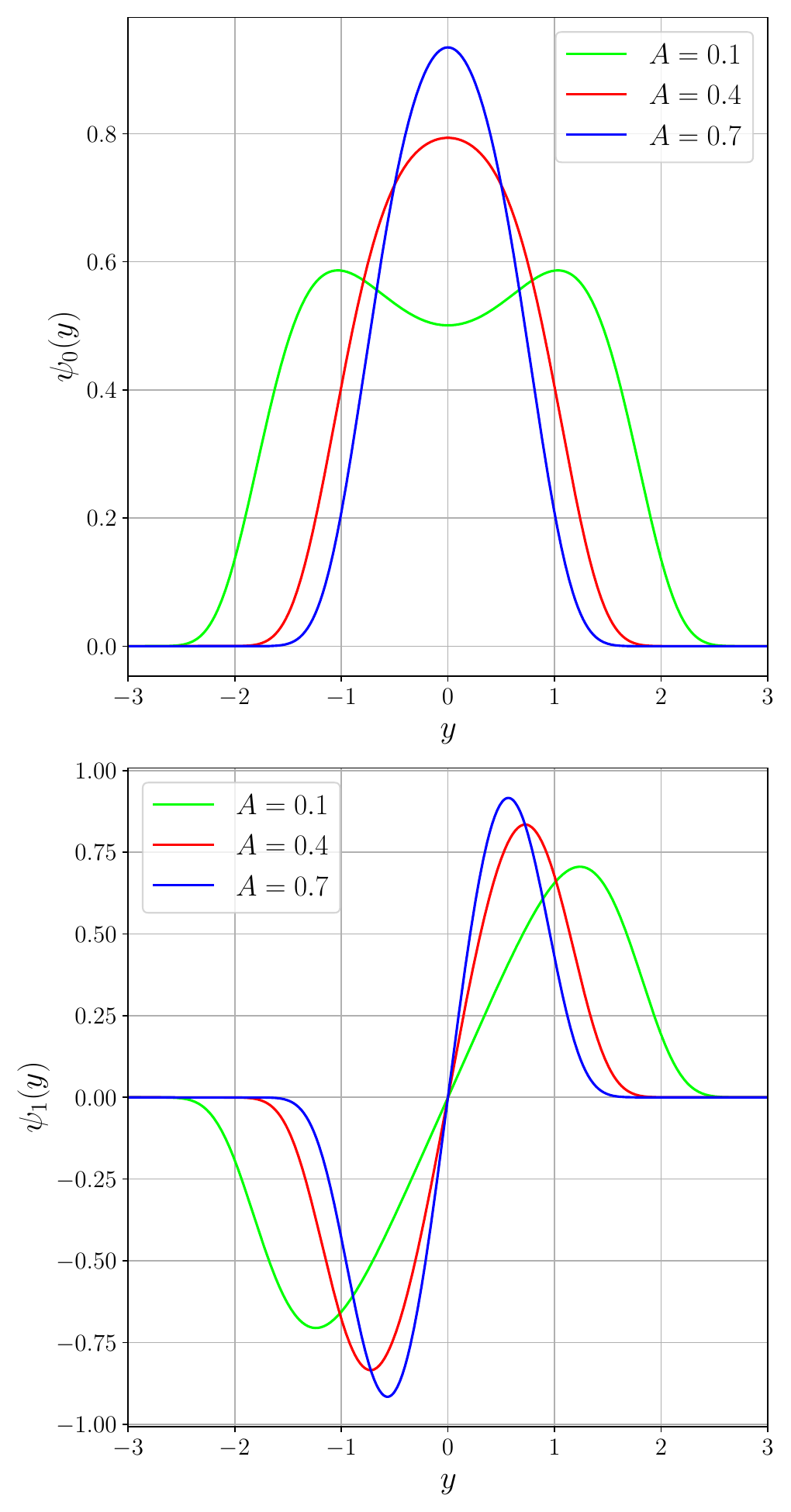}
    \caption{The profile of ground state (top) and excited state (bottom) position wavefunctions for parameter value $A=0.1,0.4$ and $0.7$.}
    \label{fig2}
\end{figure}

Figure~\ref{fig2} presents the position wavefunctions of the ground and excited states for representative values of $A$.
When $A$ is close to unity, the two potential wells are close to each other and shallow; therefore, their influence on particle localization is small.
In this regime the ground and excited states have one and two main peaks, respectively.
When $A\to0^{+}$, the two wells separate and the particle becomes localized in the two wells with comparable probability.
Consequently, the ground-state peak splits into two lobes, while the lobes of the excited state move farther apart.

\begin{figure}
    \centering
    \includegraphics[width=0.8\columnwidth]{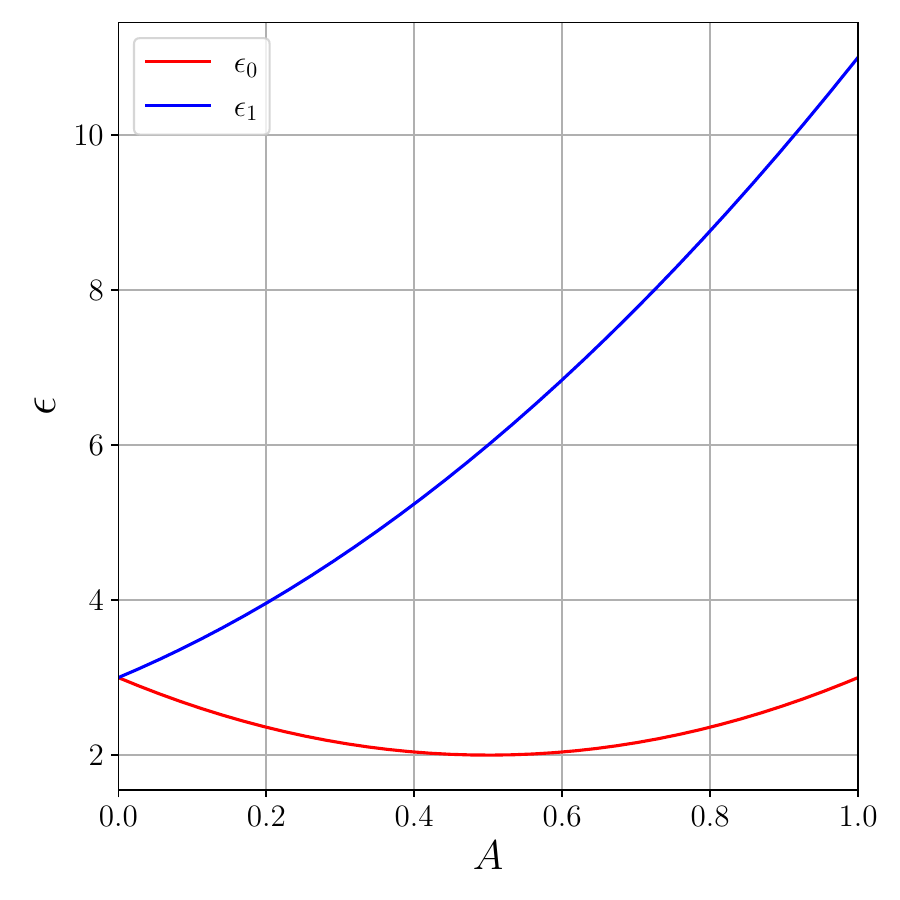}
    \caption{The energy of ground and excited state as a function of $A$.}
    \label{fig3}
\end{figure}

Figure~\ref{fig3} shows the variation of the ground- and excited-state energies as functions of $A$.
When $A\to1^{-}$, the two wells are merged and the energy levels remain clearly separated, consistent with the distinct profiles of their wavefunctions.
As $A$ decreases, the double-well structure emerges and the ground-state distribution splits so that the particle occupies both wells with comparable probability.
In the limit $A\to0^{+}$, the two levels become nearly indistinguishable because the particle is symmetrically localized in the two separated wells.
Consequently, the energy levels coincide as $A\to0^{+}$.
As $A$ decreases, $\epsilon_1$ decreases monotonically, while $\epsilon_0$ reaches a minimum at $A=0.5$.

The momentum-space wavefunctions are obtained by taking the Fourier transform of their position-space counterparts. 
The following convention is used
\begin{equation}
    \varphi_n(k) = \frac{1}{\sqrt{2\pi}}\int_{-\infty}^{+\infty} \psi_n(y)e^{-iky}\ dy.
    \label{eqn9} 
\end{equation}
The momentum coordinate $k$ is expressed in dimensionless units, consistent with the dimensionless position coordinate.
The momentum-space wavefunctions of the ground and excited states are given by 
\begin{align}
    \varphi_0(k) &= \frac{2C_0}{\sqrt{2\pi}}\int_0^{\infty} \cos(ky) \cosh(y)e^{-A\cosh(2y)}\ dy \label{eqn10} \\
    \varphi_1(k) &= -\frac{2iC_1}{\sqrt{2\pi}}\int_0^{\infty} \sin(ky) \sinh(y)e^{-A\cosh(2y)}\ dy \label{eqn11} 
\end{align}
where $C_0$ and $C_1$ are normalization coefficients of the position wavefunctions whose expression is given explicitly in equations (\ref{eqn5}) and (\ref{eqn6}), respectively. 
The integrals appearing in the above expressions can be written in terms of modified Bessel functions of the second kind, $K_\nu(z)$ with $\nu\in\mathbb{C}$; however, numerical computation is convenient for the subsequent information measures (see Appendix~\ref{appendix1}).
\begin{figure}
    \centering
    \includegraphics[width=\columnwidth]{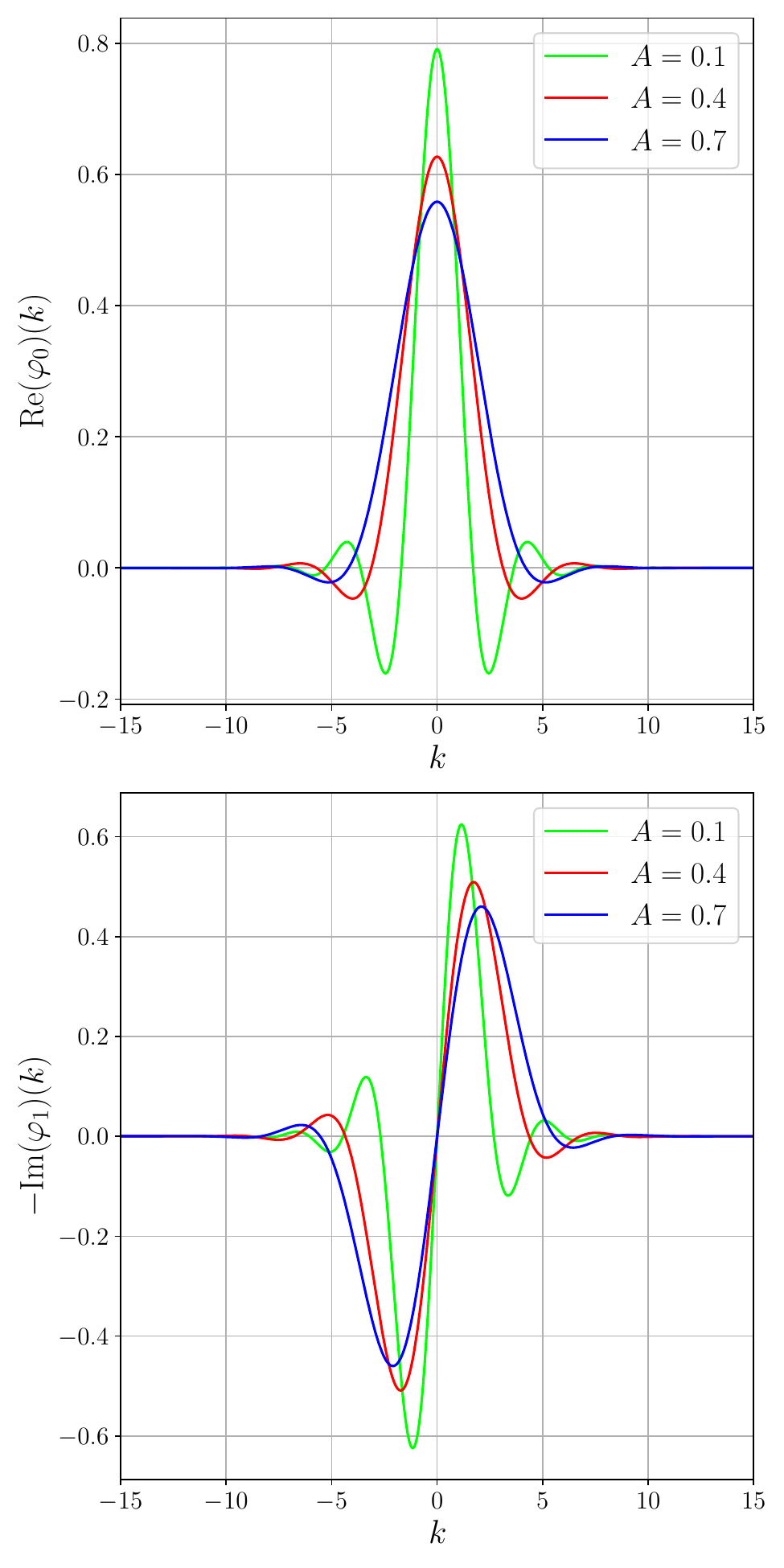}
    \caption{Momentum-space wavefunction profiles of the ground state (top) and excited state (bottom) for $A=0.1,0.4$ and $0.7$.}
    \label{fig4}
\end{figure}

Figure~\ref{fig4} shows the momentum-space wavefunctions $\text{Re}(\varphi_0)$ and $-\text{Im}(\varphi_1)$ for different values of $A$. 
These functions display oscillations that decay very rapidly. 
More oscillations are observed when $A\to0^{+}$ because the double well structure becomes more prominent.

\section{Information measures}\label{sec:qim}
\subsection{Shannon information entropy}\label{sec:shannon}
The position and momentum entropies of ground and excited states are given by 
\begin{align}
    \mathcal{S}_{\rho_0} &= -\ln(C_0^2) + C_0^2A\left[\frac{K_0(2A)}{2}+\frac{K_2(2A)}{2}+K_1(2A)\right] \nonumber \\
    &- 2C_0^2\int_0^\infty \cosh^2(y)\ln(\cosh^2(y))e^{-2A\cosh(2y)}\ dy \label{eqn15}\\
    \mathcal{S}_{\rho_1} &= -\ln(C_1^2) + C_1^2A\left[\frac{K_0(2A)}{2}+\frac{K_2(2A)}{2}-K_1(2A)\right] \nonumber \\
    &- 2C_1^2\int_0^\infty \sinh^2(y)\ln(\sinh^2(y))e^{-2A\cosh(2y)}\ dy \label{eqn16}
\end{align}
\begin{align}
    \mathcal{S}_{\gamma_0} &= -\ln\left(\frac{2C_0^2}{\pi}\right) - \frac{4C_0^2}{\pi}\int_0^\infty I_0(k) \ln I_0(k)\ dk \label{eqn17} \\ 
    \mathcal{S}_{\gamma_1} &= -\ln\left(\frac{2C_1^2}{\pi}\right) - \frac{4C_1^2}{\pi}\int_0^\infty I_1(k) \ln I_1(k)\ dk \label{eqn18}  
\end{align}
where the functions $I_j(k), j=0,1$ are given by 
\begin{align}
    I_0(k) & = \left(\int_0^\infty \cos(ky)\cosh(y)e^{-A\cosh(2y)}\ dy \right)^2 \label{eqn19} \\
    I_1(k) & = \left(\int_0^\infty \sin(ky)\sinh(y)e^{-A\cosh(2y)}\ dy \right)^2 \label{eqn20} 
\end{align}

\begin{figure}
    \centering
    \includegraphics[width=0.8\columnwidth]{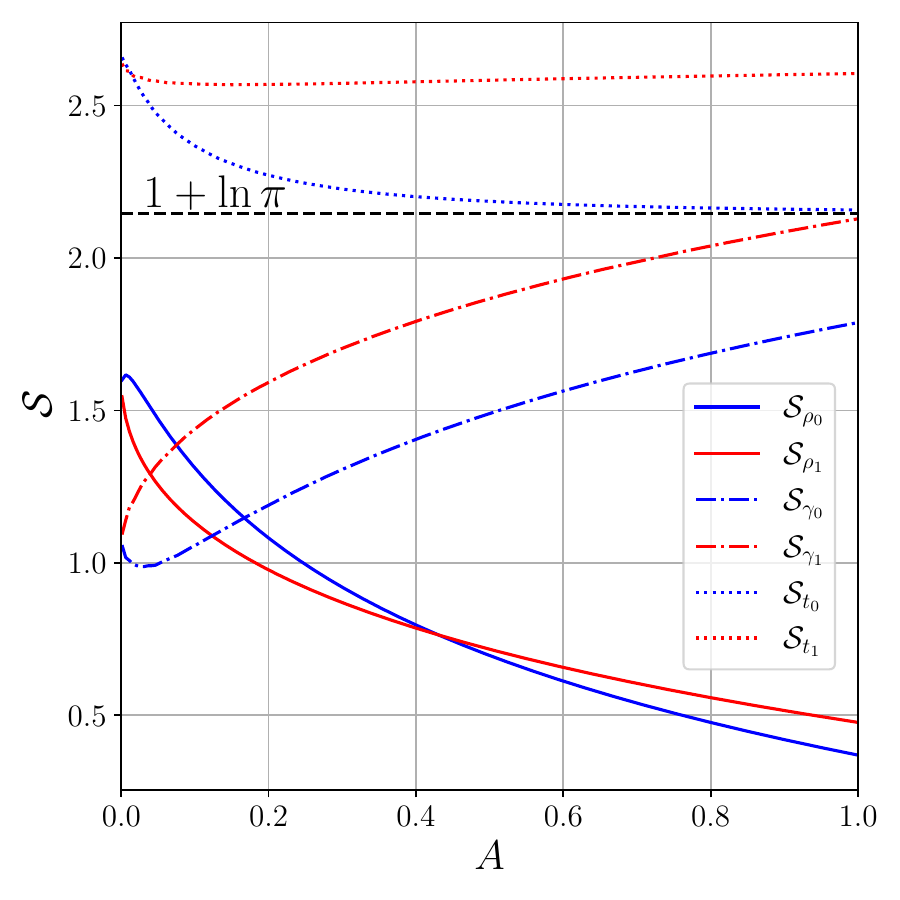}
    \caption{Position entropy (solid line), momentum entropy (dot-dashed line) and total entropy (dotted line) of the ground (blue) and excited (red) states
    as functions of $A$. The black dashed line represents the fundamental limit $1+\ln\pi$.}
    \label{fig5}
\end{figure} 

Figure~\ref{fig5} presents the Shannon entropies of the quantum states.
The position entropies of both states increase as $A\to0^{+}$ and approach the same value.
This behavior follows from the clear double-well structure at small $A$, where the particle is distributed over both wells.
As a result, the uncertainty in the particle position is high and the available position information is reduced.
As $A$ increases, the two wells begin to merge and the particle localizes near the center of the potential, reducing the position entropy.
The opposite trend is observed for the momentum entropies.
When $A$ is small, the momentum-space wavefunctions are sharply peaked around $k=0$, indicating a narrow range of likely momentum values.
As $A\to1^{-}$, the momentum distribution spreads over a broader range of $k$ values, leading to a larger momentum entropy.
Overall, the total entropy satisfies Eq.~(\ref{st_inequality}) and remains nearly constant as $A$ increases.
A key observation is that the total entropy of the ground state approaches, but does not attain, the fundamental limit.
This indicates that the ground state becomes close to Gaussian, supporting the results of Ref.~\cite{Chogle2026}.

\subsection{Onicescu energy}\label{sec:onicescu}
The Onicescu energies of the ground and excited state of the system are given by
\begin{align}
    \mathcal{O}_{\rho_0} & = C_0^4 \left[\frac{3}{8}K_0(4A) + \frac{1}{8}K_2(4A) + \frac{1}{2}K_1(4A)\right] \label{eqn24} \\
    \mathcal{O}_{\rho_1} & = C_1^4 \left[\frac{3}{8}K_0(4A) + \frac{1}{8}K_2(4A) - \frac{1}{2}K_1(4A)\right] \label{eqn25} \\
    \mathcal{O}_{\gamma_0} & = \frac{8C_0^4}{\pi^2} \int_0^{\infty} I_0^2(k)\ dk \label{eqn26} \\
    \mathcal{O}_{\gamma_1} & = \frac{8C_1^4}{\pi^2} \int_0^{\infty} I_1^2(k)\ dk \label{eqn27} 
\end{align} 

\begin{figure}
    \centering
    \includegraphics[width=0.8\columnwidth]{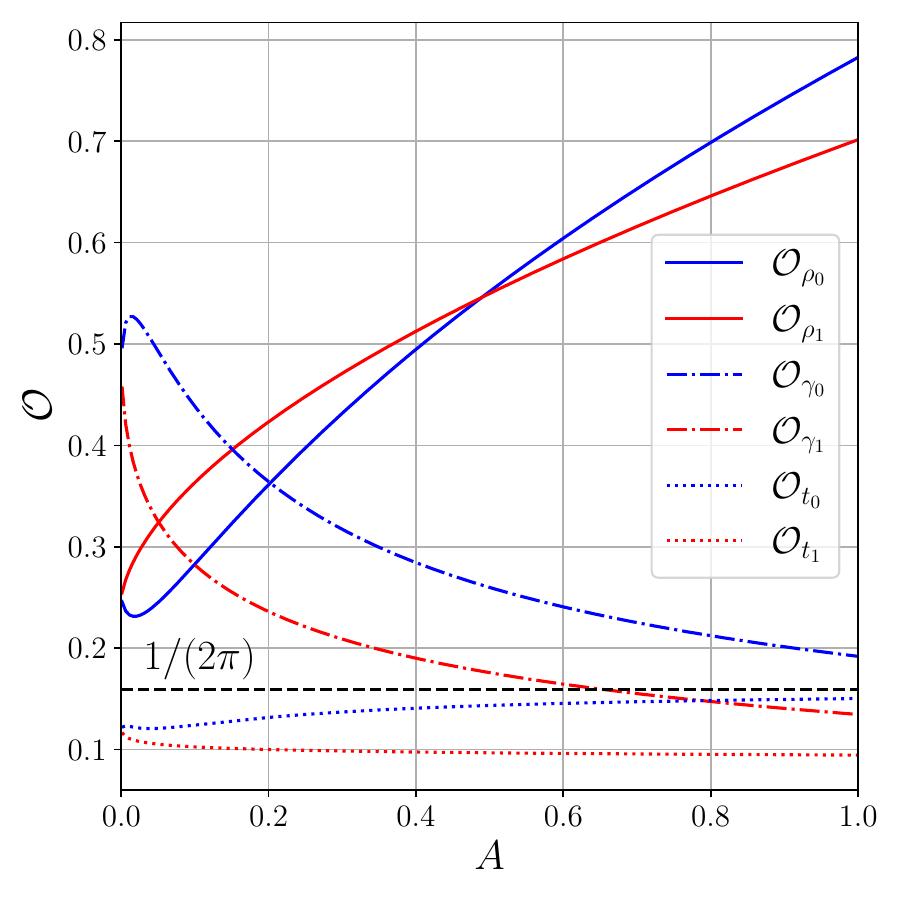}
    \caption{Onicescu energies of position (solid line), momentum (dot-dashed line) and the total product (dotted line) of the ground (blue) and excited (red) states
    as functions of $A$. The black dashed line represents the Gaussian reference value $1/(2\pi)$.}
    \label{fig6}
\end{figure} 

Figure~\ref{fig6} shows the Onicescu energies of the quantum states.
Similar to the Shannon entropies, the Onicescu energies of the two states approach the same value when $A\approx0$.
The position-space Onicescu energy increases with $A$ because the two wells merge and localize the particle near the origin, producing a more concentrated density.
Conversely, the momentum distributions show strong concentration when $A\approx0$.
Their Onicescu energies decrease as $A$ increases because the momentum-space wavefunctions spread and flatten over the $k$ axis.
For the states and parameter interval investigated here, the total product remains below the Gaussian benchmark stated in Eq.~(\ref{oni_inequality}); this numerical observation must not be interpreted as proof of a universal bound. The ground-state curve approaches $1/(2\pi)$ as $A$ increases because both conjugate densities become increasingly Gaussian-like. Thus, the apparent saturation in Fig.~\ref{fig6} is a model-specific signature of the merging-well limit, whereas the excited-state node preserves non-Gaussian structure and prevents comparable saturation.

\begin{figure}[htbp]
    \centering
    \includegraphics[width=\columnwidth]{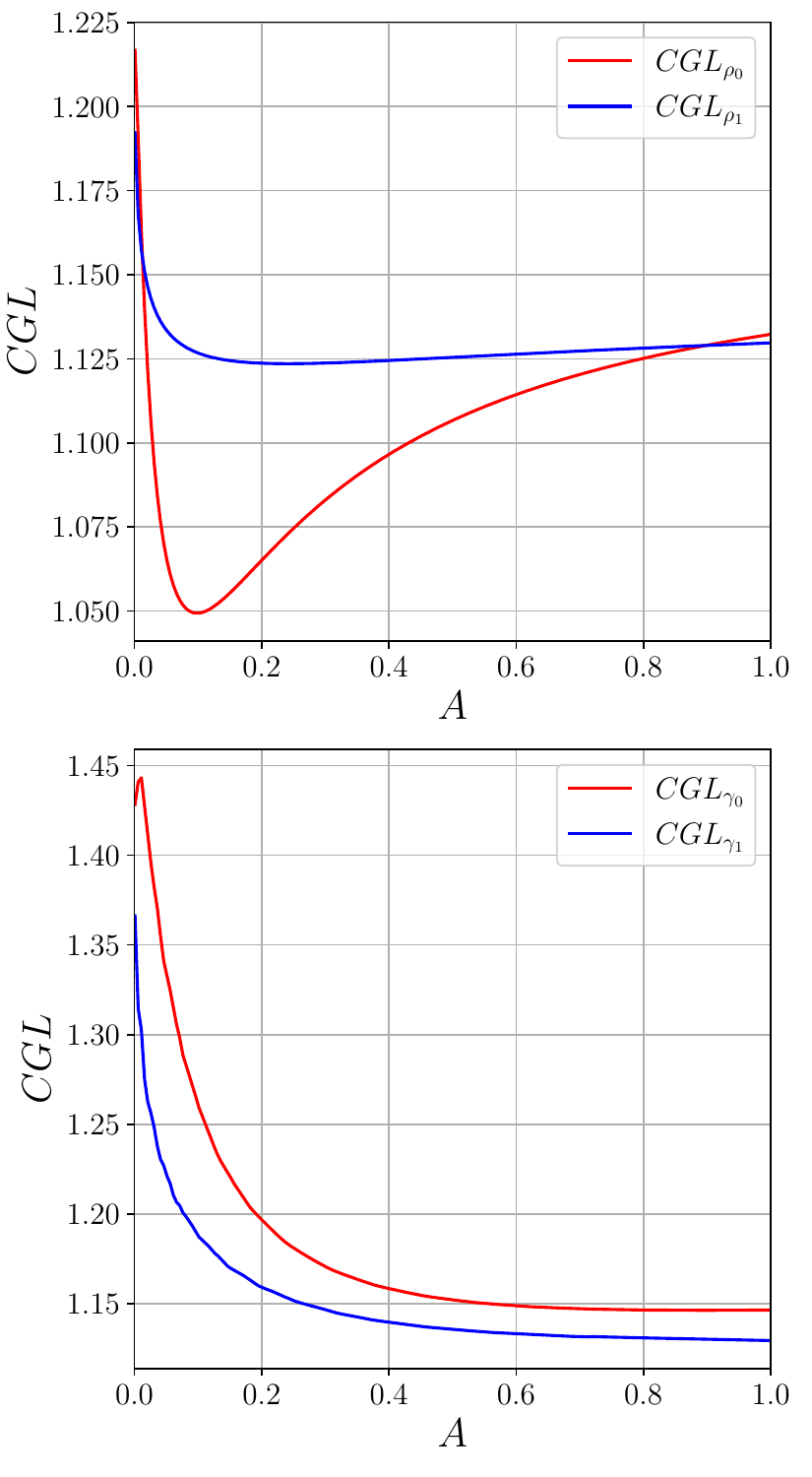}
    \caption{Statistical complexities of position (top) and momentum (bottom) waveforms as a function of $A$.}
    \label{fig6II}
\end{figure}
The statistical complexities of the position and momentum distributions of the double Morse oscillator are presented in Fig.~\ref{fig6II}.
For small $A$, the position-space complexities of the two states start close to each other because the corresponding density distributions are similar.
As $A$ increases, the two separated lobes of the ground state begin to merge and localize near the center of the well.
This behavior is reflected in the steep decrease of $\mathrm{CGL}$, followed by an increase.
By contrast, the excited-state density preserves its two-lobe structure over the same range, so its $\mathrm{CGL}$ varies more weakly.
The momentum-space complexities are larger than their position-space counterparts because of the oscillatory character of the momentum wavefunctions.
As $A$ increases, the momentum-space wavefunctions spread over momentum space, producing the monotonic trends observed in the curves.

\subsection{Fisher information}\label{sec:fisher}
The Fisher information of ground and excited states are given by 
\begin{align}
    \mathcal{I}_{\rho_0} & = C_0^2\{K_0(2A)[-4A^2+4A-2] + K_1(2A)[-2A^2+2] \nonumber \\
    & + K_2(2A)[4A^2-4A] + K_3(2A)[2A^2]\} \label{eqn30} 
\end{align}
\begin{align}
    \mathcal{I}_{\rho_1} & = C_1^2\{K_0(2A)[4A^2+4A+2] + K_1(2A)[-2A^2+2] \nonumber \\
    & + K_2(2A)[-4A^2-4A] + K_3(2A)[2A^2]\} \label{eqn31} \\
    \mathcal{I}_{\gamma_0} & = \frac{16C_0^2}{\pi} \int_0^\infty F^2_0(k)\ dk \label{eqn32} \\
    \mathcal{I}_{\gamma_1} & = \frac{16C_1^2}{\pi} \int_0^\infty F^2_1(k)\ dk \label{eqn33} 
\end{align}
where the functions $F_j(k), j=0,1$ are given by
\begin{align}
    F_0(k) & = \int_0^\infty \sin(ky)y\cosh(y)e^{-A\cosh(2y)}\ dy \label{eqn34} \\
    F_1(k) & = \int_0^\infty \cos(ky)y\sinh(y)e^{-A\cosh(2y)}\ dy \label{eqn35}. 
\end{align}

\begin{figure}
    \centering
    \includegraphics[width=0.8\columnwidth]{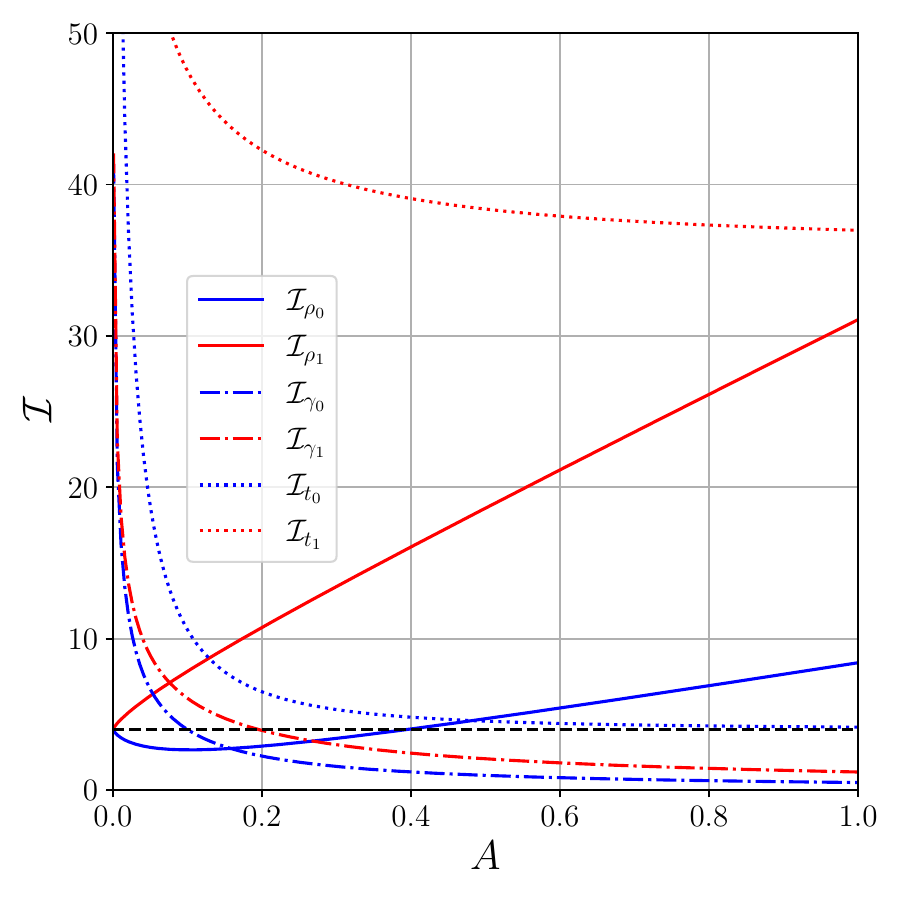}
    \caption{Fisher information of position (solid line), momentum (dot-dashed line) and the total product (dotted line) of the ground (blue) and excited (red) states
    as functions of $A$. The black dashed line represents the reference value $4$.}
    \label{fig7}
\end{figure} 

Figure~\ref{fig7} shows the Fisher information of the ground and excited states as a function of $A$.
Geometrically, a distribution with stronger local gradients has higher Fisher information than a flatter distribution.
Therefore, the momentum-space Fisher information of both states is large when $A$ is small and decreases as the distribution spreads.
Similarly, the position-space wavefunctions of both states are distributed over the two wells when $A\approx0$ and develop sharper structure near $y=0$ as $A$ increases.
The height of this central structure grows because the potential wells merge as $A\to1^{-}$ and the particle becomes localized near the origin.
The Fisher-information product satisfies Eq.~(\ref{fisher_ineq}).

\begin{figure}[htbp]
    \centering
    \includegraphics[width=\columnwidth]{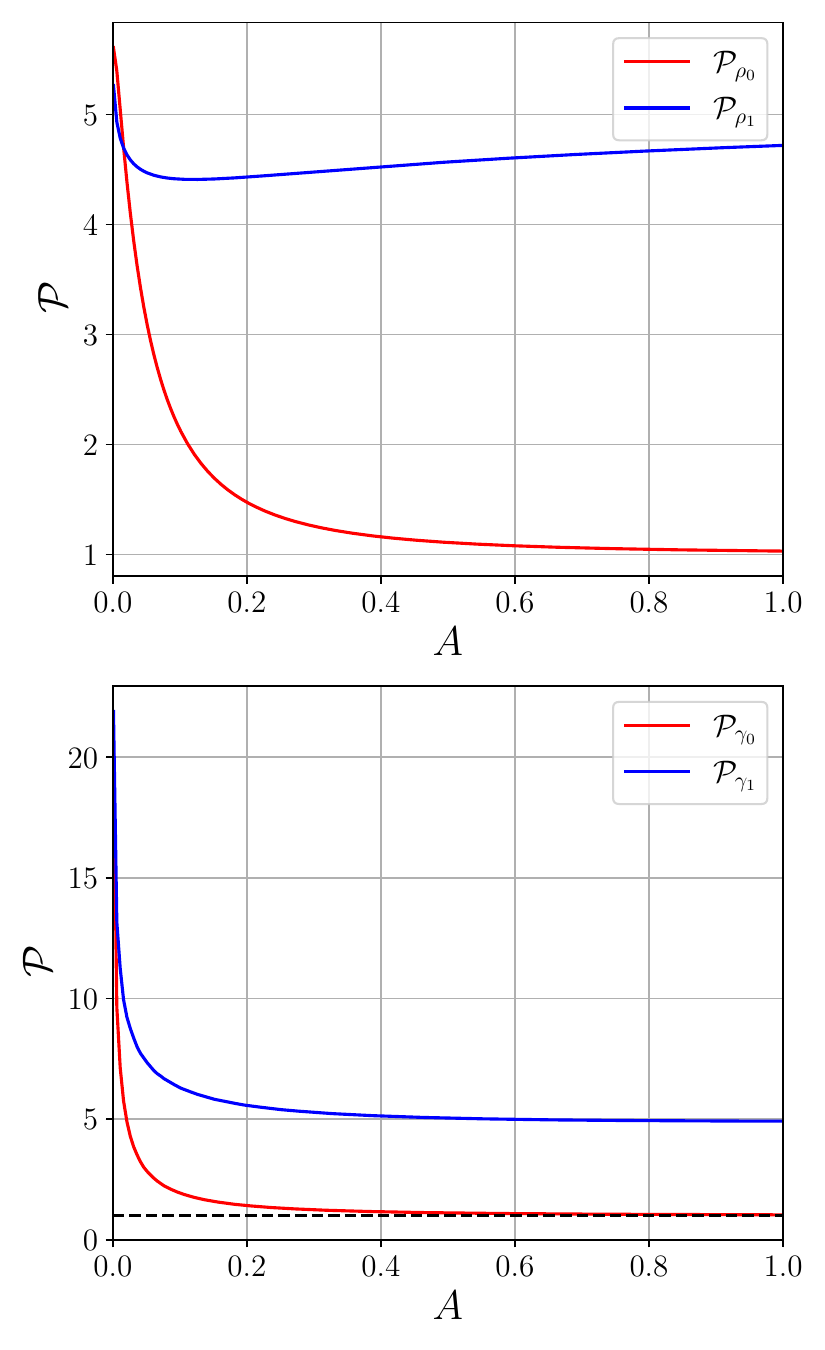}
    \caption{Fisher-Shannon product of position (top) and momentum (bottom) space distributions. The black dashed line in the bottom plot corresponds to $\mathcal{P} = 1$.}
    \label{fig7II}
\end{figure}
The Fisher--Shannon product measures the balance between global spreading (entropy power $\mathcal{J}$) and local gradient content (Fisher information $\mathcal{I}$). It satisfies the isoperimetric inequality \cite{dembo}
\begin{equation}
    \mathcal{P} \geq 1,
    \label{shannon-fisher-ineq}
\end{equation}
with equality for Gaussian distributions.
Figure~\ref{fig7II} shows the Fisher-Shannon products of the position- and momentum-space wavefunctions as functions of $A$.
The values indicate that neither distribution is Gaussian for small $A$.
They also quantify the non-Gaussian character and structural complexity of the distributions, with the momentum-space wavefunctions being more complex than their position-space counterparts.
Equivalently, the ground-state excess $\Delta_{\rm FS}=\mathcal{P}-1$ decreases toward zero as $A\to1^{-}$, quantitatively supporting the approach to Gaussian behavior.
A complementary state-level characterization of continuous-variable Gaussianity is provided by the relative-entropy measure \cite{genoni2008nong}
\begin{equation}
    \delta_\mathrm{NG}(\hat\rho) = S(\hat\rho\Vert\hat\tau)
    =S(\hat\tau)-S(\hat\rho),
    \label{eq:deltaNG}
\end{equation}
where $S$ denotes the von Neumann entropy and $\hat\tau$ is the Gaussian state with the same first moments and covariance matrix as $\hat\rho$. For the pure eigenstates considered here, $S(\hat\rho)=0$, so $\delta_\mathrm{NG}=S(\hat\tau)$. Unlike a Fisher--Shannon product computed from one marginal density, this quantity characterizes the full quantum state and can capture non-Gaussianity associated with phase-space correlations.
For the present one-mode states, we evaluate this quantity from the covariance matrix $\sigma_{ij}=\langle\{R_i-\langle R_i\rangle,R_j-\langle R_j\rangle\}\rangle/2$, with $\mathbf{R}=(y,k)$ and $[y,k]=i$. If $\nu=\sqrt{\det\sigma}$ is its symplectic eigenvalue, then
\begin{equation}
 S(\hat\tau)=g(\nu-\tfrac12),\qquad
 g(n)=(n+1)\ln(n+1)-n\ln n.
 \label{eq:gaussianEntropy}
\end{equation}
This expression specifies how the curves below were calculated from the position- and momentum-space eigenfunctions.

\begin{figure}[htbp]
    \centering
    \includegraphics[width=\columnwidth]{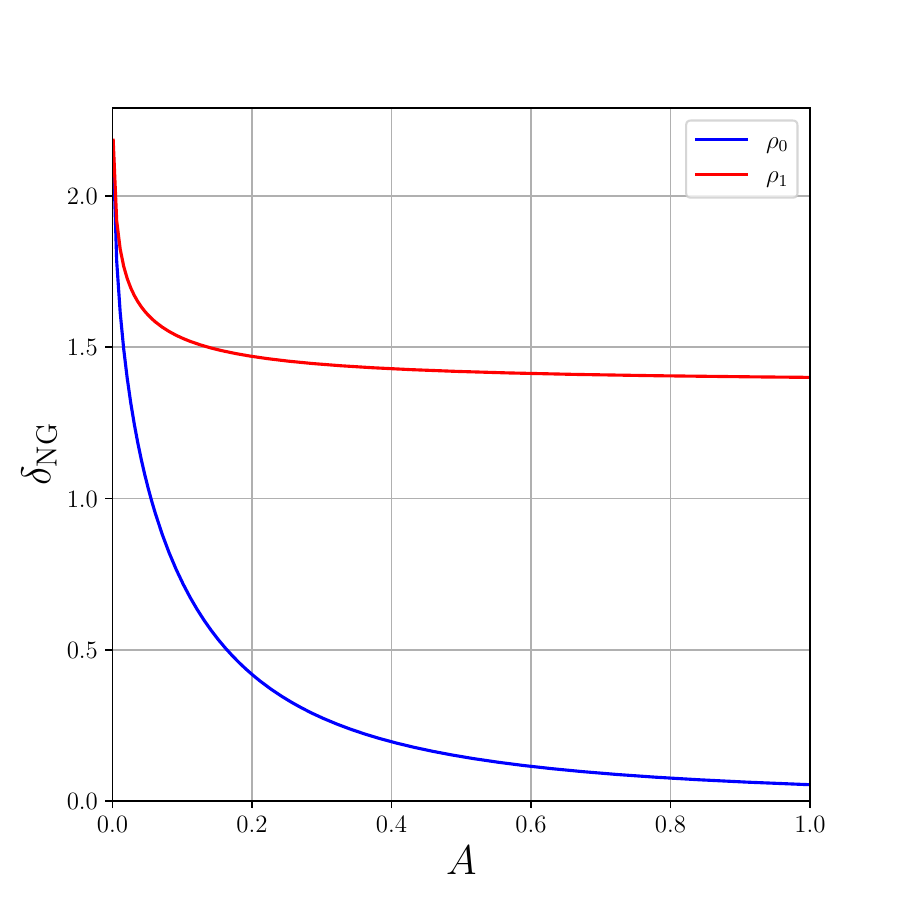}
    \caption{Relative-entropy non-Gaussianity of the ground (blue) and first excited (red) states of the double Morse oscillator as a function of $A$.}
    \label{fig:nonGaussianity}
\end{figure}

Figure~\ref{fig:nonGaussianity} shows the relative-entropy non-Gaussianity of the ground and first excited states as a function of $A$. Its decrease with $A$ is consistent with the Fisher--Shannon trends, while the larger excited-state values reflect the persistent nodal structure. Experimentally determining $\delta_\mathrm{NG}$ generally requires state tomography, or an equivalent reconstruction sufficient to determine the state entropy and covariance matrix. By contrast, each Fisher--Shannon product can be estimated from a single calibrated position- or momentum-space density, although the derivative entering Fisher information makes the estimate sensitive to noise and finite resolution. Thus, the Fisher--Shannon curves in Fig.~\ref{fig7II} provide a more directly accessible, marginal-density diagnostic, whereas $\delta_\mathrm{NG}$ provides a fuller state-level test.
\section{Conclusion}\label{sec:conc}
We studied the first two quasi-exact levels of the double Morse oscillator and analyzed their energy levels and wavefunctions in position and momentum spaces.
The relevant properties are governed by the dimensionless parameter $A\in(0,1)$, which controls the transition from two distinct wells at small $A$ to a merged single-well profile as $A\to1^{-}$.
We evaluated Shannon entropy, Onicescu energy, Fisher information, statistical complexity, and Fisher-Shannon products for both states.
These measures characterize complementary aspects of localization, uncertainty, sensitivity, and structural complexity.
For small $A$, the momentum uncertainty is lower and the momentum distribution is more sharply structured, whereas the position distribution is more delocalized over the two wells.
The opposite trend occurs as the wells merge.
As $A\to1^{-}$, the ground-state Fisher--Shannon product decreases toward unity and its Onicescu product approaches the Gaussian benchmark, providing two complementary density-based signatures of Gaussian-like behavior.
These density-based signatures are consistent with the relative-entropy non-Gaussianity of the full state, and we have outlined the different experimental information required to determine the two types of diagnostic.
The experimental interpretation discussed in Sec.~\ref{sec:experimental} further shows how the same dimensionless trends can be connected to calibrated barrier heights, well separations, local frequencies, or spectroscopy-derived potential parameters in cold-atom, trapped-ion, semiconductor, and proton-transfer platforms.
This analysis demonstrates that information-theoretic measures provide a useful diagnostic of complexity in quasi-exactly solvable double-well systems and may support future applications in quantum information and quantum-state engineering.
Because the closed-form quasi-exact sector selected here has $n=1$, our conclusions apply only to the ground and first excited states. They should not be extrapolated to higher bound states, whose additional nodes can substantially alter Fisher information, momentum oscillations, and complexity; treating those states requires numerical diagonalization or a different quasi-exact sector and is left for future work.

\section*{Acknowledgments} 
This research was funded by Khalifa University of Science and Technology through the Project ID: KU-INT-RIG-2024-8474000739 and 
was supported by KU Research Center for Advanced Intelligent Systems (AIS), Khalifa University of Science and Technology (KU-AIS).

\section*{Author Declarations}
\subsection*{Conflict of Interest}
The authors have no conflicts to disclose.

\subsection*{Ethics Approval}
The authors confirm that ethics approval was not required for this theoretical work.

\section*{Data Availability}
The data used to generate the plots in this manuscript were obtained by direct implementation of the equations and parameters provided in the text. The corresponding data can be obtained from the corresponding author upon reasonable request.

\appendix
\section{Numerical method}\label{appendix1}
The numerical procedure used to evaluate the integrals in
Secs.~\ref{sec:wave} and \ref{sec:qim} is described below.
Because the integrands of the $y$-integrals contain a
super-exponentially decaying factor, the semi-infinite integration
domain can be truncated to the finite interval
$[0,y_\mathrm{max}]$.
For a generic integrand $f(y)$, the cutoff $y_\mathrm{max}$ is chosen
such that
\begin{equation}
    \lvert f(y)\rvert \leq \epsilon
    \qquad \text{for all } y \geq y_\mathrm{max},
\end{equation}
where $\epsilon$ is a prescribed decay threshold.
For oscillatory integrands, this condition is applied to the decaying
envelope of $\lvert f(y)\rvert$.
The numerical results reported in this manuscript were obtained using
$\epsilon=10^{-6}$.
Adaptive numerical quadrature is then performed over the truncated
interval.
\bibliographystyle{unsrt}

\bibliography{references}

\end{document}